\documentclass[aps,preprint]{revtex4}%
\usepackage{amsmath}
\usepackage{graphicx}
\usepackage{amsfonts}
\usepackage{amssymb}%
\begin{document}
\title{A perturbation approach to dipolar coupling spin dynamics}
\author{G.B. Furman }
\affiliation{Physics Department, Ben Gurion University, Beer Sheva, 84105 Israel}

\begin{abstract}
A perturbation method is presented which can be applied to the description of
a wide range of physical problems that deal with dynamics of dipolarly coupled
spins in solids. The method is based on expansion $e^{\mathcal{A}+\mathcal{B}%
}$ in a perturbation series. As example of application the method, the
multile-quantum coherence dynamics in three and four spin cluster are
considered. The calculated 0Q- and 2Q intensities vs the duration of the
preparation period give closed agreement with exact results and simulations
data. The exact solutions for $J_{0Q}$ and $J_{2Q}$ coherences in four spin
system are obtained.

\end{abstract}
\pacs{33.40.+f}
\maketitle

Dipolar coupling spin dynamics represents a significant interest from both the
point of view of the general problems of physics of many bodies, and from a
position of the nuclear magnetic resonance (NMR) \cite{A.Abragam,U.Haeberlen}.
\ In solids, the evolution of a spin system under the dipole-dipole
interaction (DDI) involves many spins and leads to unsolvable problems. Even
the analysis using the numerical calculation becomes difficult because the
number of states $N=2^{n}$ is growing exponentially with the increasing of
$n$. So in existing theories, only macroscopic characteristics such as
spin-spin relaxation times, the second and the fourth moments of resonance
lines were taken into account \cite{A.Abragam}. This difficulties are very
clearly \ displayed in multiple-quantum (MQ) spin dynamics. The MQ phenomena
involves various multiple-spin transitions between the Zeeman energy levels
and form MQ coherence at times $t>\omega_{d}^{-1}$ , where $\omega_{d}$ is the
characteristic frequency of DDI \cite{J.B.Murdoch}. The problem in analytical
description of MQ is that the different modes of coherence grow in at
different times, with higher modes requiring longer excitation times than
lower modes \cite{J.Baum}. Hence, $\omega_{d}t>1$ is not a small parameter,
and, at first glance, perturbation theory methods cannot be used to study MQ
dynamics. Indeed only simple exactly solvable models of spin system such as
two and three dipolar coupling spin-1/2 \cite{A.K.Roy &
K.K.Gleason1996,S.I.Doronin2003} or one-dimensional linear chains spin
\cite{S.I.Doronin2000} system were analyzed theoretically. The last
achievement in this direction is the model with identical DDI coupling
constant for all spin pairs\cite{J.Baugh2001,M.G.Rudavets & E.B.Fel'dman2002}.
Note, that the simplified calculations are essential for the case of
indentical DDI coupling constant has been already mentioned \cite{I.J.Lowe}.
Such approaches cannot describe MQ processes except for zeroth and
second-order coherences. Thus, the development of method that can successfully
represent important features of MQ dynamics with a larger then 0Q and 2Q
coherences is needed.

Importance of the analytical description of the MQ processes in solids is that
involving the excitation of collections of dipolar coupling spins, can provide
important structural, as well as the spin dynamics informations. Moreover,
during the past few years, the NMR is considered as the best candidate
\cite{D.G.Cory1997,N.Gershenfeld,B.Kane1998} for experimental realization
quantum information processes. It was experimentally demonstrated the creating
pseudopure spin states in large clusters of coupled spins by MQ method
\cite{Khitrin} and that dynamics of the quantum entanglement is uniquely
determined by the time evolution of MQ coherences \cite{S.I.Doronin2003}.
Thus, investigation of the quantum information processes with MQ methods are
of current interest.

We present a perturbation approach to the problem of dipolar coupling spin
dynamics in solids. DDI of all spins was divided into several groups that are
characterized by the identical DDI coupling constants. Since the magnitudes of
the dipolar coupling constants vary inversely with the cube of internuclear
distance, the coupling constants are different for these groups.

Our main idea is to take into account in MQ NMR dynamics influence of the
groups with different degree of accuracy. Spin groups with smaller DDI
coupling constants can be considered as perturbation (spins located far apart)
while the nearest neighbours are taken into account exactly. As the result, we
can develop a perturbation method that allows obtaining the description of the
MQ with a large coherence evolution under DDI in an analytical form. The
proposed approach will be a powerful method to describe wide ranges of
physical problems that deal with dynamics of dipolar coupled spins in solid.
On the one hand, this approach uses the advantages of exactly solvable models
\cite{M.G.Rudavets & E.B.Fel'dman2002,A.R.Kessel2002}. On the other hand, it
simplifies calculations by using a perturbation technique. Results will show
that the perturbation method can be applied to solve complex spin-dynamics
problem and to obtain the solution in an analytical form. The method is based
on the differential method \cite{W.Magnus,R.M.Wilcox1967} expresses
$e^{\mathcal{A}+\mathcal{B}}$ as an infinite product of exponential operators
\cite{R.M.Wilcox1967}. In the case when the norm of operator $\mathcal{B}$ is
small then one of operator $\mathcal{A}$, $\left\Vert \mathcal{B}\right\Vert
<\left\Vert \mathcal{A}\right\Vert $, we will try to obtain the perturbation
series which takes into account up to second order terms in ratio $\left\Vert
\mathcal{B}\right\Vert /\left\Vert \mathcal{A}\right\Vert $. Then the problems
in description the MQ dynamics will be significantly simplified.

Let us consider a spin system with Hamiltonian, $\mathcal{H}$ which includes
only two parts with different DDI constants $\alpha$ and $\beta$:
$\mathcal{H}=\mathcal{A}+\mathcal{B}$ , where $\mathcal{A}=\alpha A$ and
$\mathcal{B}=\beta B$, $\alpha=\left\Vert \mathcal{A}\right\Vert $ and
$\beta=\left\Vert \mathcal{B}\right\Vert $ are the norms of operators
$\mathcal{A}$ and $\mathcal{B}$, respectively ($\alpha>$ $\beta$ and $\left[
A,B\right]  \neq0$). The evolution of the spin system is governed by
propagator
\begin{equation}
e^{-it\mathcal{H}}=e^{-it\left(  \alpha A+\beta B\right)  }. \tag{1}%
\end{equation}
We seek to express (1) as a series in such that%

\begin{equation}
e^{-it\left(  \alpha A+\beta B\right)  }=e^{-it\beta B}\sigma\left(  t\right)
, \tag{2}%
\end{equation}
where operator $\sigma\left(  t\right)  $ obeys the differential equation
\cite{R.M.Wilcox1967,Bellman}%

\begin{equation}
i\frac{d\sigma\left(  t\right)  }{dt}=\alpha A^{\left(  0\right)  }\left(
t\right)  \sigma\left(  t\right)  , \tag{3}%
\end{equation}
with initial condition
\begin{equation}
\sigma\left(  0\right)  =1 \tag{4}%
\end{equation}
and $A^{\left(  0\right)  }\left(  t\right)  =e^{-it\beta B}Ae^{it\beta B}$.
Assume that $\alpha t\geq1$ and $\beta t<1$. First, we will restrict ourself
by keeping only first-order terms that are linearly proportion to $\beta$.
This leads to
\begin{equation}
i\frac{d\sigma^{\left(  0\right)  }\left(  t\right)  }{dt}=\alpha\left(
A-it\beta\left[  B,A\right]  \right)  \sigma^{\left(  0\right)  }\left(
t\right)  . \tag{5}%
\end{equation}
To solve Eq.(5) we will use the iterative method. We will search the solution
of Eq.(5) as a series on parameter $\left(  \alpha t\right)  ^{n}$ $:$%

\begin{equation}
\sigma_{A}\left(  t\right)  =\sum_{n=0}^{\infty}\sigma_{A}^{\left(  n\right)
}\left(  t\right)  \tag{6}%
\end{equation}

Taking into account that $\sigma_{A}^{\left(  0\right)  }\left(  t\right)  =1$
the solution of Eq.(5) for $n=1$ can be obtained:%

\begin{equation}
\sigma_{A}^{\left(  1\right)  }\left(  t\right)  =\left(  -i\alpha t\right)
A. \tag{7}%
\end{equation}
For $n=2$, ones obtains
\begin{equation}
\sigma_{A}^{\left(  2\right)  }\left(  t\right)  =\frac{\left(  -i\alpha
t\right)  ^{2}}{2}\left(  A^{2}+\frac{\beta}{\alpha}\left[  B,A\right]
\right)  \tag{8}%
\end{equation}
Keeping only linear in $\frac{\beta}{\alpha}$ terms the following expression
for operator $\sigma_{A}^{\left(  m\right)  }\left(  t\right)  $ can be obtained%

\begin{equation}
\sigma_{A}^{\left(  m\right)  }\left(  y\right)  =\frac{\left(  -i\alpha
t\right)  ^{m}}{m!}\left(  A^{m}-\frac{\beta}{\alpha}\left(  \left(
m-1\right)  BA^{m-1}-\sum_{j=0}^{m-2}A^{m-1-j}BA^{j}\right)  \right)  \tag{9}%
\end{equation}
Using Eq. (9) we obtain%
\begin{equation}
e^{-it\left(  \alpha A+\beta B\right)  }=\left(  1-\beta\sum_{n=0}^{N}%
\alpha^{n}\frac{\left(  it\right)  ^{n+1}}{\left(  n+1\right)  !}\sum
_{j=0}^{n}\frac{\left(  -1\right)  ^{j}\left(  n\right)  !}{j!\left(
n-j\right)  !}A^{j}BA^{n-j}\right)  e^{-i\left(  \alpha tA+\frac{\beta}%
{\alpha}\frac{\left(  it\alpha\right)  ^{N+2}}{\left(  N+2\right)  !}\left[
B,A\right]  _{N+1}\right)  } \tag{10}%
\end{equation}
where $\left[  B,A\right]  _{N+1}$ denotes the repeated commutators $\left[
[[...\left[  [B,A],A\right]  ...A\right]  _{N+1}$ . After summation over $j$
in Eq. (10) we have%

\begin{equation}
e^{-it\left(  \alpha A+\beta B\right)  }=\left(  1-\frac{\beta}{\alpha}%
\sum_{n=0}^{N}\frac{\left(  i\alpha t\right)  ^{n+1}}{\left(  n+1\right)
!}\left\{  B,A^{n}\right\}  \right)  e^{-i\left(  \alpha tA+\frac{\beta
}{\alpha}\frac{\left(  it\alpha\right)  ^{N+2}}{\left(  N+2\right)  !}\left\{
B,A^{N+2}\right\}  \right)  }, \tag{11}%
\end{equation}
where
\begin{equation}
\left\{  B,A^{0}\right\}  =B\ \text{and }\left\{  B,A^{n+1}\right\}  =\left[
\left\{  B,A^{n}\right\}  ,A\right]  .\text{\ \ \ } \tag{12}%
\end{equation}
In the limit as the number of steps $N\rightarrow\infty$ we obtain that
$\lim_{N=\infty}\left(  \frac{\beta}{\alpha}\frac{\left(  it\alpha\right)
^{N+2}}{\left(  N+2\right)  !}\right)  =\allowbreak0$ . Consequently, the
exponent in (11) can be presented in the limit as $N\rightarrow\infty$ in the
following form: $\lim_{N=\infty}e^{-i\left(  \alpha tA+\frac{\beta}{\alpha
}\frac{\left(  it\alpha\right)  ^{N+2}}{\left(  N+2\right)  !}\left[
B,A\right]  _{N+1}\right)  }=e^{-i\alpha tA}$ , which does not include any
terms with $B$, and the summing over $n$ up to indefinite, results in%
\begin{equation}
e^{-it\left(  \alpha A+\beta B\right)  }=\left(  1-i\frac{\beta}{\alpha}%
\int_{0}^{\alpha t}dxe^{-ixA}Be^{ixA}\right)  e^{-i\alpha tA}, \tag{13}%
\end{equation}
which is a well known formula for expansion an exponential operator in a
perturbation series \cite{Bellman}. To obtain expansion containing only linear
to $\frac{\beta}{\alpha}$ terms we have to require that $\frac{\beta}{\alpha
}\frac{\left(  t\alpha\right)  ^{N+2}}{\left(  N+2\right)  !}\ll1$. This
requirement imposes restrictions also on time: $t\ll\frac{1}{\alpha}\left(
\frac{\alpha}{\beta}\left(  N+2\right)  !\right)  ^{\frac{1}{N+2}}$. So for
the smallest of times $t\ll\frac{1}{\alpha}\left(  \frac{\alpha}{\beta}\left(
N+2\right)  !\right)  ^{\frac{1}{N+2}}$ or \ \ $t\ll\frac{1}{\beta}$, the Eq.
(10) includes only terms linear in $\beta$ . In an analog way, we obtain the
expansion up to second order in the ratio $\frac{\beta}{\alpha}:$%
\begin{align}
&  e^{-it\left(  \alpha A+\beta B\right)  }=\{1-\frac{\beta}{\alpha}\sum
_{m=0}^{\infty}\sum_{k=0}^{\infty}[\frac{\left(  -1\right)  ^{m}\left(
ix\right)  ^{k+m+1}}{m!k!\left(  k+m+1\right)  }A^{m}BA^{k}\nonumber\\
&  +\left(  \frac{\beta}{\alpha}\right)  ^{2}\sum_{l=0}^{\infty}\sum
_{p=0}^{\infty}\frac{\left(  -1\right)  ^{m+p}(ix)^{k+l+m+p+2}}%
{m!k!l!p!\left(  l+p+1\right)  }\frac{A^{l+m}BA^{p}BA^{k}}{\left(
k+l+m+p+2\right)  }]\}e^{-ixA} \tag{14}%
\end{align}

The series expansion (14) can be used not only for the small parameter
$\frac{\beta}{\alpha}<1$, but independently, for the parameter $x=\alpha t$.
Formula (14) can be easily generalized for a case when the exponential
operator contains arbitrary number of the non-commutative operators and can be
extended to include various power of the operators. Eqs. (14) appears to be
complex at glance, but in fact it is quite simple to use, as the following
examples will illustrate.

Let us consider a cluster of three dipolar-coupled spin-$\frac{1}{2}$ nuclei.
The MQ dynamics in the rotating frame is described by propagator (1), where
the time-independent average Hamiltonian is given by
\begin{equation}
\mathcal{H}=-\frac{1}{2}\sum_{j<k}d_{jk}\left(  I_{j}^{+}I_{k}^{+}+I_{j}%
^{-}I_{k}^{-}\right)  \tag{15}%
\end{equation}
and $I_{j}^{+}$ and $I_{j}^{-}$ are the raising and lowering operators for
spin $j$. The dipolar coupling constant, $d_{jk}$, for any pair of nuclei $j$
and $k$ in the cluster, is given by
\begin{equation}
d_{jk}=\frac{\gamma^{2}\hbar}{2r_{jk}^{3}}\left(  1-3\cos\theta_{jk}\right)  ,
\tag{16}%
\end{equation}
where $\gamma$ is the gyromagnetic ratio of the nuclei, $r_{jk}$ is the
internuclear spacing, and $\theta_{jk}$ is the angle the vector \ $\vec
{r}_{jk}$\ makes\ with the external magnetic field. In the high-temperature
approximation the density matrix at the end of the preparation period is given
by
\begin{equation}
\rho\left(  t\right)  =e^{-i\mathcal{H}t}\rho\left(  0\right)  e^{i\mathcal{H}%
t} \tag{17}%
\end{equation}
where $\rho\left(  0\right)  $ is the initial density matrix in the
high-temperature approximation
\begin{equation}
\rho\left(  0\right)  =\sum_{j=1}^{3}I_{j}^{z}, \tag{18}%
\end{equation}
$I_{j}^{z}$ is the projection of the angular momentum operator on the
direction of the external field for an spin $j$. The average Hamiltonian (15)
can be divided into the three parts according to the number of the different
coupling constant $d_{12}>d_{23}>d_{13}$ :
\begin{equation}
H=H_{12}+H_{23}+H_{13}, \tag{19}%
\end{equation}
where
\begin{equation}
H_{jk}=-\frac{d_{jk}}{2}\left(  I_{j}^{+}I_{k}^{+}+I_{j}^{-}I_{k}^{-}\right)
\text{ with }j\neq k\text{ and }j,k=1,2,3. \tag{20}%
\end{equation}
The experimentally observed values are the intensities, $J_{nQ}\left(
t\right)  $ of multiple-quantum coherences:
\begin{equation}
J_{nQ}\left(  t\right)  =\frac{1}{Tr\rho^{2}\left(  0\right)  }\sum
_{p,q}\text{ }\rho_{pq}^{2}\left(  t\right)  \text{ for }n=m_{zp}-m_{zq},
\tag{21}%
\end{equation}
where $m_{zp}$ and $m_{zq}$ are the eigenvalue of the initial density matrix
(18). The perturbation method described above is used to calculate the time
evolution of MQ coherences. Using expansion (14) with $\alpha A=H_{12}$ and
$\beta B=H_{23}+H_{13}$ and keeping terms up to eighth order in $x=\alpha t$ ,
the normalized $0$-quantum ($J_{0Q}$) and $2$-quantum ($J_{2Q}$) intensities
are given by
\begin{align}
J_{0Q}  &  =1-\frac{8x^{2}}{3}+\frac{32x^{4}}{9}-\frac{256x^{6}}{125}%
+\frac{512x^{8}}{945}\nonumber\\
&  +\left(  \frac{\beta}{\alpha}\right)  ^{2}\left(  \frac{8x^{2}}{3}%
+\frac{40x^{4}}{9}-\frac{176x^{6}}{45}+\frac{1544x^{8}}{945}\right)  \tag{22}%
\end{align}
and%
\begin{align}
J_{2Q}  &  =-\frac{4x^{2}}{3}+\frac{16x^{4}}{9}-\frac{128x^{6}}{135}%
+\frac{256x^{8}}{945}\nonumber\\
&  -\left(  \frac{\beta}{\alpha}\right)  ^{2}\left(  \frac{4x^{2}}{3}%
+\frac{32x^{4}}{9}-\frac{128x^{6}}{45}+\frac{1024x^{8}}{945}\right)  \tag{23}%
\end{align}
where $\left(  \frac{\beta}{\alpha}\right)  ^{2}=\left(  \frac{d_{23}}{d_{12}%
}\right)  ^{2}+\left(  \frac{d_{13}}{d_{12}}\right)  ^{2}$. Let us compare
formulas (22) and (23) with results from Eq. (14) which the terms with
$x=\alpha t$ will be taken into account exactly. By summing over $n$, $m$,
$l$, and $k$ up to indefinitely in (14), we obtain the analytical expressions
of the intensities of $0$ - quantum%
\begin{equation}
J_{0Q}=\frac{1}{3}\left\{  \cos4x-2\left[  \left(  \frac{\beta}{\alpha
}\right)  ^{2}-1\right]  +2\left(  \frac{\beta}{\alpha}\right)  ^{2}\left(
\cos x+\cos3x-\cos4x-x\sin4x\right)  \right\}  \tag{24}%
\end{equation}
and of $2$-- quantum
\begin{equation}
J_{2Q}=-\frac{\sin2x}{3}\left\{  2x\left(  \frac{\beta}{\alpha}\right)
^{2}\cos2x+2\left[  \left(  \frac{\beta}{\alpha}\right)  ^{2}-\left[  \left(
\frac{\beta}{\alpha}\right)  ^{2}-1\right]  \cos x\right]  \sin x\right\}
\tag{25}%
\end{equation}
Now let us compare intensities (22) - (25) with the exact solution
\cite{A.K.Roy & K.K.Gleason1996}. Figs. 1 and 2 show the evolution of the
normalized 0Q and 2Q coherences for three spin cluster, where $\frac{\beta
}{\alpha}=0.3$ and at $t=0$ the spin system is in thermal equilibrium (18).
All approaches, perturbations (Eqs. (22) -(25)) and exact \cite{A.K.Roy &
K.K.Gleason1996}, give closed agreement up to $x=0.75$ (in unit of $\frac
{1}{\alpha}$), both for $0Q$ - and $2Q$-- coherences. The exact account of
influence of the nearest neighbours gives a good agreements up to $x=2$.

As a second example we consider is cluster consists of four spin arrange in
corners of square in an external magnetic field penpendicular to the square
plane. In this case the MQ spin dynamics is described by the average
Hamiltonian
\begin{equation}
H=H_{1}+H_{2}, \tag{26}%
\end{equation}
with two different dipolar coupling constants $D_{1}$ and $D_{2}$ , where
$D_{1}$ and $D_{2}$ are the dipolar coupling constants between nearest
neighbors and spins at opposite sites, respectively ($\frac{\beta}{\alpha
}=\frac{D_{2}}{D_{1}}=\frac{1}{2\sqrt{2}}$) where%

\begin{equation}
H_{1}=\left(  -\frac{D_{1}}{2}\right)  \sum_{j=1}^{4}\left(  I_{j}^{+}%
I_{j+1}^{+}+I_{j}^{-}I_{j+1}^{-}\right)  =\alpha A \tag{27}%
\end{equation}
and
\begin{equation}
H_{2}=\left(  -\frac{D_{2}}{2}\right)  \sum_{j=1}^{2}\left(  I_{j}^{+}%
I_{j+2}^{+}+I_{j}^{-}I_{j+2}^{-}\right)  =\beta B \tag{28}%
\end{equation}
Using the expansion (14) up to eighth order in $x=\alpha t$, the normalized
$0$-quantum ($J_{0Q}$)%
\begin{equation}
J_{0Q}=1-2x^{2}+\frac{7}{4}x^{4}-\frac{13}{18}x^{6}+\frac{5}{28}x^{8}-\left(
\frac{\beta}{\alpha}\right)  ^{2}\left(  x^{2}-\frac{13}{6}x^{4}+\frac
{121}{60}x^{6}-\frac{599}{630}x^{8}\right)  \tag{29}%
\end{equation}
and $2$-quantum ($J_{2Q}$)%

\begin{equation}
J_{2Q}=-\frac{x^{2}}{4}+\frac{x^{4}}{4}-\frac{1}{9}x^{6}+\frac{1}{35}%
x^{8}-\left(  \frac{\beta}{\alpha}\right)  ^{2}\left(  \frac{x^{2}}{8}%
-\frac{x^{4}}{3}+\frac{11x^{6}}{30}-\frac{58x^{8}}{315}\right)  \tag{30}%
\end{equation}
intensities can be determined. Formulas (29) and (30) will be compare with
results from Eq.(14) in which the terms describing interaction of the
neighbour spins will be taken into account exactly. By summation over $n$,
$m$, $l$, and $k$ up to indefinitely in Eq.(14) we obtain the analytical
expressions of the intensities of $0$ - quantum%

\begin{equation}
J_{0Q}=\frac{1}{4}\left(  1+\sin^{2}2x+2\sin^{2}\sqrt{2}x\right)  +\frac
{x^{2}}{8}\left(  \frac{\beta}{\alpha}\right)  ^{2}\left(  2\sin^{2}%
2x-\frac{\sqrt{2}}{x}\sin2\sqrt{2}x\right)  \tag{31}%
\end{equation}
and of $2$-- quantum%
\begin{equation}
J_{2Q}=-\frac{1}{8}\left(  \sin^{2}2x+2\sin^{2}\sqrt{2}x\right)  +\frac{x^{2}%
}{16}\left(  \frac{\beta}{\alpha}\right)  ^{2}\left(  2\sin^{2}2x-\frac
{\sqrt{2}}{x}\sin2\sqrt{2}x\right)  \tag{32}%
\end{equation}
coherences$.$

To control the perturbation results (29) - (32) we obtained the exact solution
for $J_{0Q}$%

\begin{equation}
J_{0Q}=\frac{9}{4}-\frac{1}{2}\cos\left(  2x\sqrt{2+\left(  \frac{\beta
}{\alpha}\right)  ^{2}}\right)  +\frac{1}{4}\cos4x\cos\left(  2x\frac{\beta
}{\alpha}\right)  -\cos^{2}\left(  x\frac{\beta}{\alpha}\right)  \tag{33}%
\end{equation}
and for $J_{2Q}$%
\begin{equation}
J_{2Q}=-\frac{3}{4}+\frac{1}{4}\cos\left(  2x\sqrt{2+\left(  \frac{\beta
}{\alpha}\right)  ^{2}}\right)  +\frac{1}{4}\sin^{2}2x\cos\left(
2x\frac{\beta}{\alpha}\right)  +\frac{1}{2}\cos^{2}\left(  x\frac{\beta
}{\alpha}\right)  \tag{34}%
\end{equation}
coherences and fulfilled the numerical analysis of the MQ dynamics$.$ The
exact solutions (33) and (34) and computer simulation of the MQ coherences of
four spins cluster have been obtained with a PC using the MATLAB package.

Figs. 3 and 4 show, that perturbation results (29) and (30) are in the
agrement with (31) and (32) and the exact solutions (33) and (34) and with the
simulation data one up to $x=1$ (in unit of $\frac{1}{\alpha}$). Calculations
in which the interaction between the nearest neighbours is taken into account
exactly (Eqs (31) and (32)) are in close agreement with the exact solutions
(Eqs.(33) and (34)) and simulation data up to $x=3$.$\allowbreak$

In conclusion, a perturbation method was developed which is based on the
expansion of operator exponent in a perturbation series. Then the perturbation
approach was applied to the description the MQ spin dynamics in solids. The
analytical expressions for 0Q and 2Q dynamics in a three and four spin
clusters in solids were obtained. In the four spin cluster the exact solution
was obtained. The calculated 0Q- and 2Q intensities vs the duration of the
preparation period agree well with exact solutions for three \cite{A.K.Roy &
K.K.Gleason1996} and for four spin clusters (Eqs.(33) and (34)).

The developed method can be extended to include various power of the operators
with small norm and applied to the description widely range of physical
problems deal with dynamics of dipolar coupling spins in solids. The results
in an analytical form can be use to extract from experimental data the dipolar
constants and the molecular structure information.

The authors are grateful to E. Fel'dman (Institute of Problems of Chemical
Physics, Chernogolovka) and A. Khitrin (Kent State University, Kent) for
useful discussions. V. Meerovich , V. Sokolovsky and S.William (Ben-Gurion
University, Beer Sheva) for assistance in carrying out computing calculation.
This research was supported by a Grant from the U.S.-Israel Binational Science
Foundation (BSF).

\bigskip

\bigskip

Captions for figures.

Figure 1

Fig.1 Time dependences (in units of $\frac{1}{\alpha}$ ) of the normalized
intensities of 0Q coherence (solid-line is exact solution \cite{A.K.Roy1996},
dot-line is the calculation using Eq.(22) and dash-line is the calculation
using Eq.(24))

Figure 2

Fig.2 Time dependences (in units of $\frac{1}{\alpha}$ ) of the normalized
intensities of 2Q coherence (solid-line is exact solution\cite{A.K.Roy1996},
dot--line is the calculation using Eq.(23) and dash-line is the calculation
using Eq.(25))

Figure 3

Fig. 3 Time dependences (in units of $\frac{1}{\alpha}$ ) of the normalized
intensities of 0Q coherence in four spin cluster. Solid-line is exact solution
(Eq.(33), dot-line is the calculation using Eq.(29), dash-line is the
calculation using Eq.(31), and open circle is computer simulations.

Figure 4

Fig. 4 Time dependences (in units of $\frac{1}{\alpha}$ ) of the normalized
intensities of 2Q coherence in four spin cluster. Solid-line is exact solution
(Eq.(34), dot-line is the calculation using Eq.(30), dash-line is the
calculation using Eq.(32), and open circle is computer simulations.

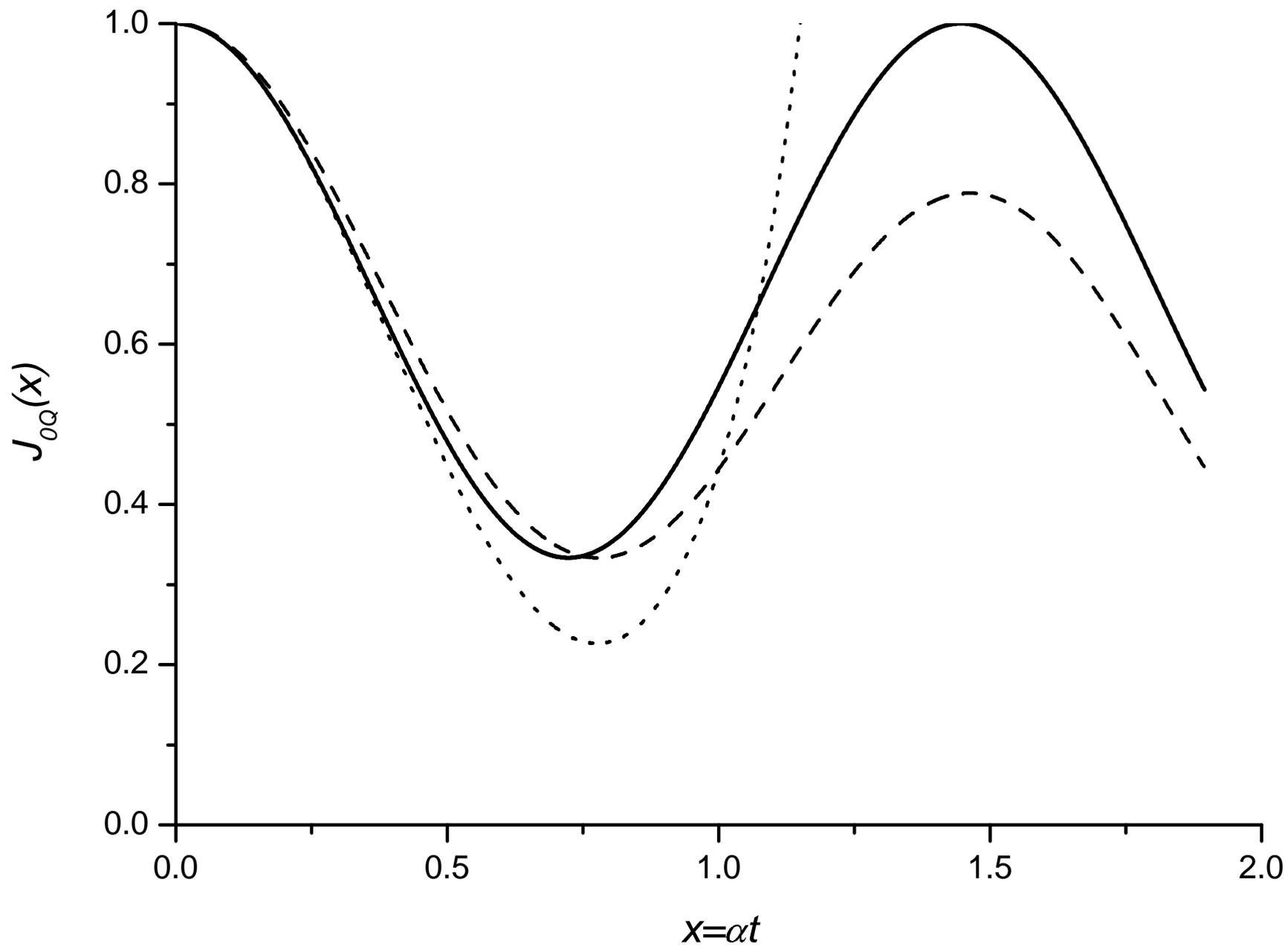
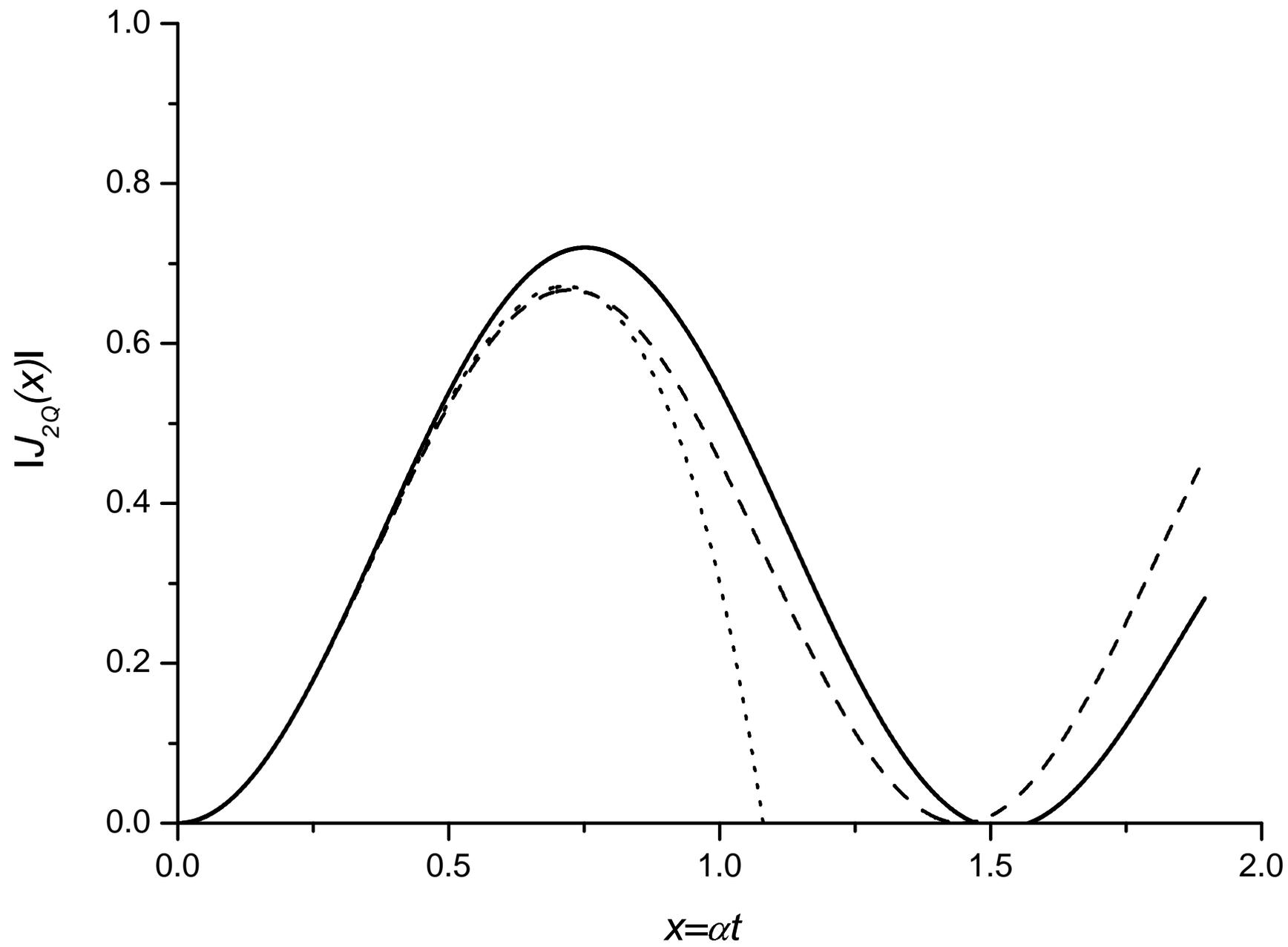
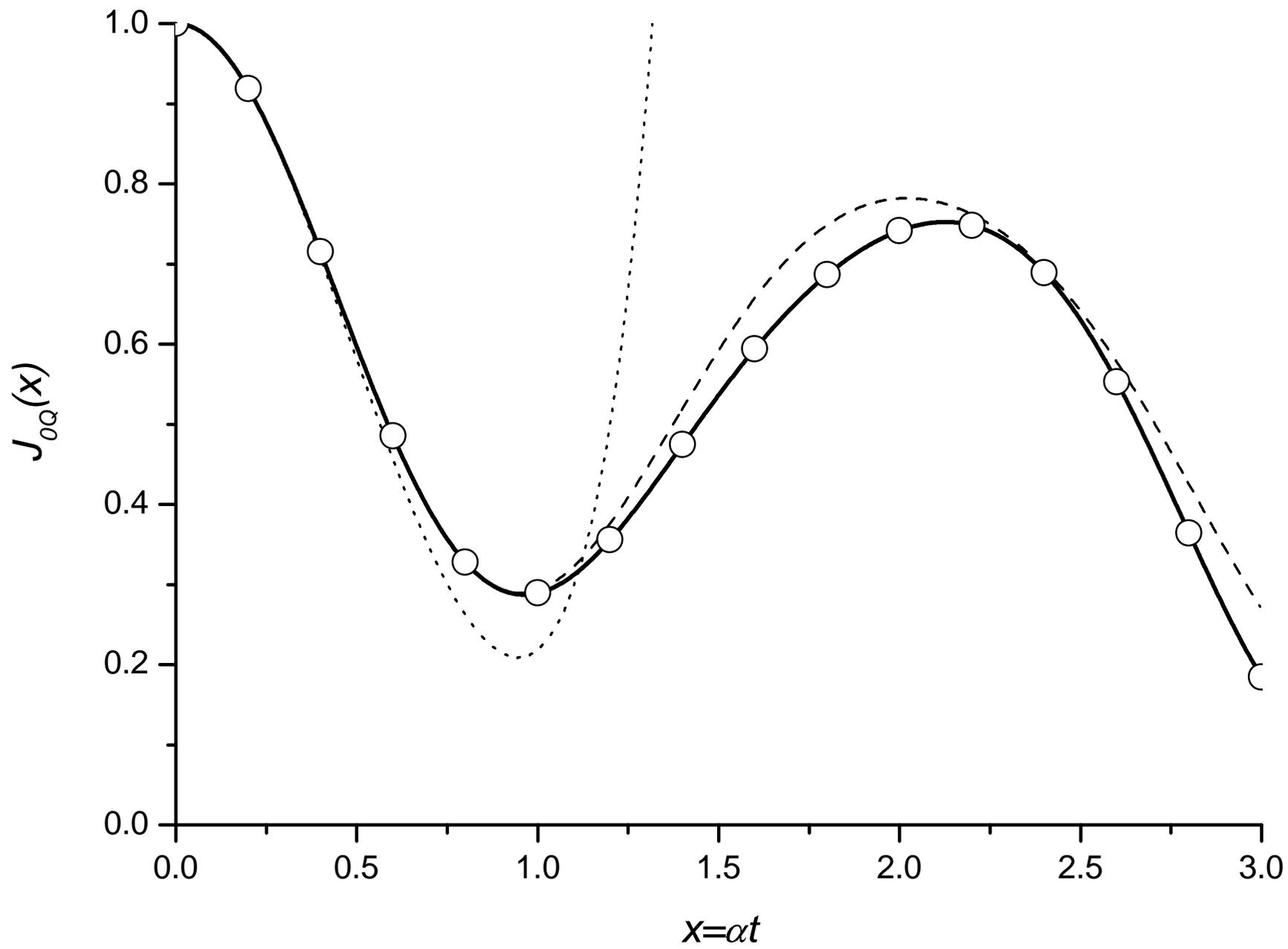
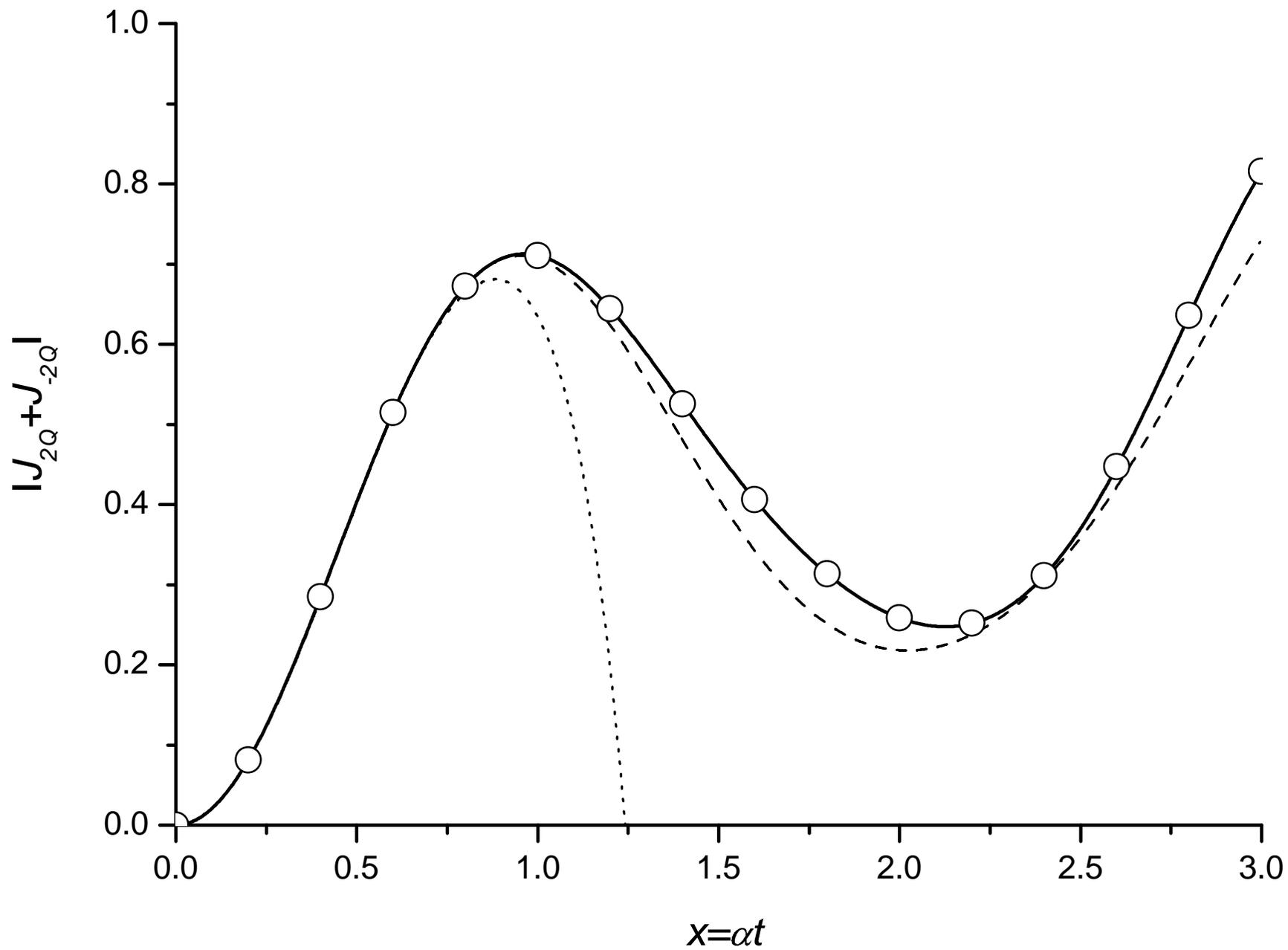
\end{document}